\documentclass[aps,prb,nobibnotes,showpacs,amsfonts,amsmath,amssymb,twocolumn,letterpaper]{revtex4}
\usepackage{epsfig}\usepackage{amsmath}
\usepackage{amsmath,amsfonts,amssymb,graphics,graphicx,epsfig,color,times,bbm}

\begin{document}

\title{Kerr nonlinearities and nonclassical states with superconducting qubits and nanomechanical resonators}
\author{F. L. Semi\~ao}
\affiliation{Departamento de F\'isica, Universidade Estadual de Ponta Grossa - Campus Uvaranas, 84030-900 Ponta Grossa, Paran\'a, Brazil}
\author{K. Furuya}
\affiliation{Institute of Physics ``Gleb Wataghin'', P.O. Box 6165, University of Campinas - UNICAMP, 13083-970 Campinas, SP, Brazil}
\author{G. J. Milburn}
\affiliation{School of Physical Sciences - University of Queensland, Brisbane, Queensland 4072, Australia}

\begin{abstract}
We propose the use of a superconducting charge qubit capacitively coupled to two resonant nanomechanical resonators to generate Yurke-Stoler states, i.e. quantum superpositions of pairs of distinguishable coherent states 180$^\circ$ out of phase with each other. This is achieved by effectively implementing Kerr nonlinearities induced through application of a strong external driving field in one of the resonators. A simple study of the effect of dissipation on our scheme is also presented, and lower bounds of fidelity and purity of the generated state are calculated. Our procedure to implement a Kerr nonlinearity in this system may be used for high precision measurements in nanomechanical resonators.
\end{abstract}



\maketitle
\section{Introduction}
 Quantum nonlinear dynamics is an important topic in physics. In quantum optics, nonlinear interactions have been widely used to generate nonclassical field states, such as squeezed or sub-Poissonian light \cite{sspl}. A special class of optical nonlinearity results in an intensity dependent phase shift,  commonly known as the Kerr effect. In the single mode case, the time evolution of an initial coherent state, under the influence of such a Kerr medium and very low loss, will evolve into a quantum superposition of two coherent states 180$^\circ$ out of phase with each other. This was first discovered by Yurke and Stoler \cite{ys}, and since then such states have been called Yurke-Stoler states. A single-mode Kerr medium preserves the photon statistics but modifies the quadrature uncertainties generally leading to squeezing \cite{sspl}. 

There is great interest in observing this quantum nonlinear couplings in solid state systems. This would allows us to deepen our current understanding of the classical-quantum frontier by studying how long can superpositions of mesoscopically distinct states survive in such systems. Some interesting proposals involving nanomechanical resonators have been published during the last years. In one scheme \cite{jacobs}, the use of a time dependent drive in a Cooper pair box coupled to a nanomechanical resonator is shown to generate a number of nonlinear Hamiltonians for the latter. By parametrically driving a nanomechanical resonator capacitively coupled to a superconducting coplanar waveguide one can generate interesting nonlinear Hamiltonians suitable for generation and detection of squeezed states as proposed in \cite{woolley}. In this system, entangled states in temperatures up to tens of milliKelvin may be achieved as discussed in \cite{vitali}. Nanomechanical oscillators have also been shown to be feasible for coupling to other important physical systems besides Cooper pair boxes or microwave fields of coplanar wave guides. Nanomechanical resonators may, for instance, be coupled to Bose-Einstein condensates \cite{Treutlein}, trapped ions \cite{tian,ions} or spin degrees of freedom of a sample of neutral atoms in the gas phase \cite{wang}.

In this paper, we propose a theoretical scheme to engineer Kerr Hamiltonians using a system composed of a Cooper pair box capacitively coupled to two {\sl resonant} nanomechanical resonators.  We show in Sec. \ref{sec2} that such nonlinear Hamiltonians can be achieved in a dispersive regime by appropriately choosing the system's parameters and  by using a properly tuned strong classical field in one of the resonators. The integration of superconducting qubits with nanoresonators is an important topic and has been previously considered in \cite{cleland,schwab,irish,irish2,guo,solano}. We start from a well known Hamiltonian describing the interaction between a charge qubit and two resonant nanomechanical resonators in a quantum regime \cite{irish,schwab1,huang,bose07},  and we then include an external driving in one of the oscillators. By considering the regime of intense driving, we show that a nonlinear Kerr-type effective Hamiltonian  may be obtained. This Hamiltonian is induced by the common coupling of the resonators with the qubit and intense external driving. This is the central result of this paper, and as an application, we show in Sec. \ref{sec3} how to generate the Yurke-Stoler state in the normal modes of the nanomechanical resonators. We also discuss the zero temperature decoherence in a particular regime of relaxation, and evaluate both the fidelity and the purity of the generated superposition state. Finally, we would like to point out that the ability to implement Kerr nonlinearities in nanomechanical resonators has recently been shown to find applications also in high precision measurements. In a recent paper \cite{woolley2}, Woolley {\sl et al.} have proposed a new protocol for high precision measurement in a nanomechanical resonator that makes explicit use of such nonlinearities. This might be a potential application for the results presented in this paper. In Sec. \ref{sec4} we draw some conclusions.
\section{The model and the Kerr type interactions}
\label{sec2}
\begin{figure}[h]
\begin{center}
\includegraphics[scale=0.4]{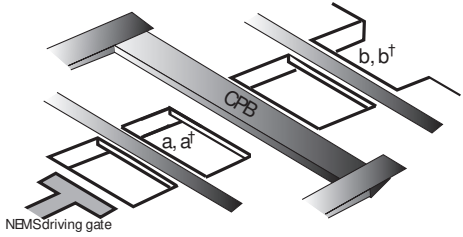}
\end{center}
\caption{Two nanomechanical resonators (with lowering operators $a$ and $b$) are capacitively coupled to a Cooper pair box (CPB). One of the oscillators is driven by a classical force.}
\label{fig1}
\end{figure}

The simplest charge qubit, the Cooper pair box (CPB), consists of a small superconducting island with an excess number, $n$, of Cooper-pairs, connected by a tunnel junction (capacitance $C_J$ and Josephson coupling $E_J$) to a superconducting electrode. External control is achieved by the application of a voltage gate $V_g$ coupled to the CPB via a gate capacitor with capacitance $C_g$.  More details can be found in the review \cite{review}. For specific qubit proposals and decoherence analysis see \cite{squbit}.  In our study, we will assume that the CPB is coupled capacitively to two nano-eletromechanical systems (NEMS) \cite{bose07}, as depicted in Fig.(\ref{fig1}). 
In the two level approximation for the CPB, the capacitive coupling between the qubit and two NEMS is described by the Hamiltonian \cite{bose07}
\begin{eqnarray}\label{jth}
H & = & \hbar\omega a^ \dagger a+\hbar\omega b^ \dagger b+\frac{\hbar \omega_0 }{2}\bar{\sigma}_z+\frac{\hbar \bar{\Delta}}{2}\bar{\sigma}_x+ \hbar\lambda_1(a+a^\dagger)\bar{\sigma}_z \nonumber\\
& & +\hbar\lambda_2(b+b^\dagger)\bar{\sigma}_z+\hbar g(a e^{i\omega_e t}+a^\dagger  e^{-i\omega_e t}),\label{oH}
\end{eqnarray}
where $a,a ^\dagger$ are the raising and lowering operators for the driven NEMS \cite{ruskov}, $b,b^\dagger$ are the raising and lowering operators for the other NEMS with the same resonance frequency $\omega$, and $g$ represents the amplitude of the external nanomechanical drive (frequency $\omega_{\rm{e}}$). The parameters appearing in (\ref{oH}) are given by 
\begin{eqnarray}
\hbar\omega_{\rm{0}}  &=& -4E_c(1-2n_g), \\
\hbar \bar{\Delta}  &=&  - 2E_J\cos(\pi\phi/\phi_0), \\
\hbar\lambda_i  &=&e\frac{V_g C_{g,i}}{C_\Sigma d_i}\sqrt{\frac{\hbar}{2 m \omega}},
\end{eqnarray}
where $C_{g,i}$ is the capacitance between the CPB and $i$-th nanomechanical bias gate, $C_\Sigma$ is the total capacitance, $d_i$ is the distance between the $i$-th nanomechanical bias gate and the CPB, and $m$ is the mass of the NEMS. The couplings can be made different by varying, for example, the distances $d_i$ or applying DC voltages to the resonators \cite{schwab06}. Our goal now is to show how the application of the external driving  field may be used to engineer nonclassical states.  We first make a rotation of the qubit to new variables $\bar{\sigma}_\alpha\rightarrow \sigma_\alpha$: 
\begin{eqnarray}\label{h1}
 H & = & \hbar\omega a^\dagger a+\hbar\omega  b^\dagger b+\hbar g(a e^{i\omega_{\rm{e}} t}+a^\dagger e^{-i\omega_{\rm{e}} t})+ \nonumber\\
& &  \frac{\hbar\bar{\Omega}}{2}\sigma_z+
\hbar\lambda_1 (a+a^\dagger)(\cos\theta \sigma_z-\sin\theta \sigma_x)
\nonumber\\
 & & +\hbar\lambda_2 (b+b^\dagger)(\cos\theta \sigma_z-\sin\theta \sigma_x),
 \end{eqnarray}
where 
\begin{eqnarray}\label{sdef}
\cos\theta &=& \frac{\omega_{\rm{0}}}{\bar{\Omega}},\\
\sin\theta  &=& \frac{\bar{\Delta}}{\bar{\Omega}},\\
\bar{\Omega}  &=& (\omega_{\rm{0}}^2+\bar{\Delta}^2)^{1/2}.
\end{eqnarray}
Now, moving to a rotating frame with frequency $\omega_{\rm{e}}$, and setting $\bar{\Omega}=\omega_{\rm{e}}$ and $\delta =\omega -\omega_{\rm{e}}$, we get
\begin{eqnarray}\label{h2}
 H & = & \hbar\delta a^\dagger a+  \hbar\delta b^\dagger b +
\hbar g(a+a^\dagger)+ \hbar\lambda_1(a e^{-i\omega_e t}+  \nonumber\\
& & a^\dagger e^{i\omega_e t}) [\cos\theta \sigma_z-\sin\theta ( \sigma_+ e^{i\omega_e t}+\sigma_-e^{-i\omega_e t})] +  \nonumber\\ 
& & \hbar\lambda_2 (b e^{-i\omega_e t} +b^\dagger e^{i\omega_e t}) [\cos\theta \sigma_z-\sin\theta ( \sigma_+ e^{i\omega_e t}\nonumber\\
& & +\sigma_-e^{-i\omega_e t})].
 \end{eqnarray}
We can now make the rotating wave approximation, to get the interaction picture Hamiltonian,
\begin{eqnarray}
 H & = & \hbar g(a e^{-i\delta t}+a^\dagger e^{i\delta t})-\hbar\lambda_1 \sin\theta (a \sigma_+e^{-i\delta t}+a^\dagger \sigma_-e^{i\delta t})\nonumber\\ & & \mbox{} -\hbar\lambda_2 \sin\theta (b \sigma_+e^{-i\delta t} +b^\dagger \sigma_-e^{i\delta t})
 \label{Hrwa}
 \end{eqnarray}
An interesting situation appears when one takes the dispersive approximation 
 ($|\delta|\gg\lambda_1,\lambda_2,g$) for the above Hamiltonian (applying similar methods to those described in \cite{onetwo}). 
In this regime, the Hamiltonian (\ref{Hrwa}) may be approximated by
\begin{eqnarray}\label{hext}
H=\hbar\Omega a^\dagger a \sigma_z+\hbar \chi b^\dagger b \sigma_z+\hbar\Delta \sigma_x+\hbar r(a^\dagger b+ab^\dagger)\sigma_z,
\end{eqnarray}
where 
\begin{eqnarray}\label{nrel}
\Omega &=&-\frac{\lambda_1^2}{\delta}\sin^2\theta,\\ 
\chi &=&-\frac{\lambda_2^2}{\delta}\sin^2\theta,\\
\Delta &=&\frac{g\lambda_1}{\delta}\sin\theta, \\
r&=&-\frac{\lambda_1 \lambda_2}{\delta}\sin^2\theta.
\end{eqnarray}
The  Hamiltonian (\ref{hext}) can be diagonalized by using new bosonic composite operators
$a_1=(\cos{\frac{\gamma}{2}}a+\sin{\frac{\gamma}{2}}b)$ and $a_2=(-\sin{\frac{\gamma}{2}}a+\cos{\frac{\gamma}{2}}b)$ with appropriate choice for $\gamma$.
We set from now on $\Omega=\chi$, i.e. $\lambda_1=\pm\lambda_2$, 
 since for this case the simple choice  $\gamma= \frac{\pi}{2}$ solves the problem. In terms of the new operators
$a_1=2^{-1/2}(a+b)$ and $a_2=2^{-1/2}(a-b)$, $H$ is written  (setting $\hbar=1$) as
\begin{equation}
\tilde{H}_{+(-)} = \xi a^\dag_{1(2)}a_{1(2)}\sigma_z +\Delta
\sigma_x
\end{equation}
where $\zeta=-\frac{2\lambda_1^2}{\delta}\sin^{2} \theta$. We will now show that in the regime  $|\Delta| \gg |\zeta|$, a {\sl Kerr type} Hamiltonian can be generated.
From $\zeta=-\frac{2\lambda_1^2}{\delta}\sin^{2} \theta $, we see that $|\Delta|\gg |\zeta|$ implies that we must have $g\gg 2\lambda_i \sin \theta  $ ($i=1\,{\rm{or}}\,2$),
 i.e a strong driving $(g\gg \lambda_i)$. To make this clear, lets us assume $\lambda_1=\lambda_2$  and $\Delta>0$. By transforming $H_+$ to an interaction picture with respect to $\Delta\sigma_x$, one obtains
\begin{eqnarray}
\tilde{\mathcal{V}}_+(t)=\frac{\zeta}{2}\{a_1^\dag a_1[(\sigma_z-i\sigma_y)e^{2i\Delta t}+(\sigma_z+i\sigma_y)e^{-2i\Delta t}]\}.\nonumber\\
\end{eqnarray} 
Now, if one defines the operator $A=a_1^\dag a_1 (\sigma_z-i\sigma_y)$ and the constant $\lambda=\frac{\zeta}{2}$, the above Hamiltonian will read $\tilde{\mathcal{V}}_+(t)=\lambda(Ae^{i2\Delta t}+A^\dag e^{-i2\Delta t})$. It can be shown \cite{onetwo} that for $\Delta>>\lambda$, the  effective Hamiltonian $\tilde{\mathcal{V}}_{+}^{\,{\rm{eff}}}=\hbar\frac{\lambda^2}{2\Delta}[A,A^\dag]$ can be used. By evaluating this commutator, one finds
\begin{eqnarray}
\tilde{\mathcal{V}}_{+}^{\,{\rm{eff}}}=\mu(a_1^\dag a_1)^2\sigma_x\label{K}
\end{eqnarray}
where $\mu=\zeta^2/2\Delta$. Remarkably, this Hamiltonian mimics the single mode Kerr effect. If the CPB is prepared in an eigenstate of $\sigma_x$, the bosonic mode will follow a decoupled evolution under the nonlinear Hamiltonian $\mu(a_1^\dag a_1)^2$. Going back to the definitions, one can see that the magnitude of the nonlinearity $\mu$ is in fact controlled by the system parameters $\lambda_1$ (coupling constant for the interaction of resonator $a$ with the qubit), $g$ (related to the amplitude of the classical driving), and $\delta$ (detuning between driving field and nanoresonators). 
 Thus, it is possible to control the the intensity of the present Kerr 
type effect, which is always important in the applications.
\section{Yurke-Stoler state and inclusion of dissipation in the NEMS}
\label{sec3}
Consider now the initial preparation, $|\psi(0)\rangle=|\alpha\rangle_a|\alpha\rangle_b|+\rangle_x$, i.e both resonators in coherent states with the same amplitude $\alpha$, and the CBP in an eigenstate of $\sigma_x$ with eigenvalue equal to one. In the transformed space of the composite modes $a_1$ and $a_2$, this initial state becomes $|\tilde\psi(0)\rangle=|\sqrt{2}\alpha\rangle_1|0\rangle_2|+\rangle_x$. It means that the composite mode-$1$ is initially in a coherent state $|\alpha_1=\sqrt{2}\alpha\rangle_1$, mode-$2$ in the vacuum state $|\alpha_2=0\rangle_2$, and the qubit in the eigenstate of $\sigma_x$ corresponding to the eigenvalue $1$. For this initial condition, Hamiltonian (\ref{K}) leads to the following time evolved state:
\begin{eqnarray}\label{ki}
|\tilde{\psi}_I(t)\rangle=\left[e^{-|\alpha_1|^2/2}\sum_{n=0}^{\infty}\frac{(\alpha_1)^n}{\sqrt{n!}}e^{-it\mu n^2}|n\rangle_1\right]|0\rangle_2|+\rangle_x,\nonumber\\
\end{eqnarray}
with $\alpha_1=\sqrt{2}\alpha$. For an interaction time $t_I$ such that $\mu t_I=\pi/2$, the state (\ref{ki}) evolves to
$
|\tilde{\psi}_I(t_I)\rangle=|{\rm{YS}}\rangle_1|0\rangle_2|+\rangle,
$
where
\begin{eqnarray}
|{\rm{YS}}\rangle_1=\frac{|\alpha_1\rangle_1+i|-\alpha_1\rangle_1}{\sqrt{2}}
\end{eqnarray}
is the Yurke-Stoler state. We remark that no measurement whatsoever was needed to generate this state, so this scheme is deterministic. If initially one prepares \linebreak $|\psi(0)\rangle=|\alpha\rangle_a|-\alpha\rangle_b|+\rangle_x$ and choose $\lambda_1=-\lambda_2$, a Yurke-Stoler state is generated in the mode-$2$. Many applications for superpositions of coherent states have been suggested in the quantum optics and quantum information literature \cite{applic}, along with a considerable variety of generation protocols \cite{cats}.

Since we have performed a perturbation approach of the problem (effective Hamiltonians), it is now important to make a brief discussion about the experimental values of the parameters and the feasibility of the regimes we used. From (\ref{h2}) to (\ref{Hrwa}), we have realized a rotating wave approximation, and this is justified when $\lambda_1,\lambda_2\ll\omega,\bar{\Omega},g$. According to experimental reference \cite{huang}, it is currently possible to achieve $\omega/2\pi=1.0$ GHz. For charge qubits, ordinary values for $\bar{\Omega}$ are also about a few gigahertz \cite{nat2}. In principle, the external driving $g$ may also be of the same order or even stronger than $\omega$ and $\bar{\Delta}$. We have also demanded $\bar{\Omega}=\omega_e$, and this means that the frequency of the drive field is also of a few gigahertz. Taking all these into account, we see that the coupling constants $\lambda_1$ and $\lambda_2$ must be at most around a few megahertz for the rotating wave approximation to be valid. This seems not be a problem since such coupling constants may be tuned by changing the distance between the CPB and the nanoresonators or through additional DC voltages on the resonator.  When going from (\ref{Hrwa}) to (\ref{hext}), we took the dispersive regime that demands $\lambda_1,\lambda_2\ll\delta$. Again, this might not be a problem since $\lambda_i$ depends on $d_i$. Finally, our last approximation corresponds to the regime of strong driving $\lambda_1,\lambda_2 \ll g$. This seems to be easy to achieve since $g$ is externally controlled via a driving gate and do not depend on the fabrication features of the CPB or the resonators. 
 
It is well known that superposition states of this kind are easily corrupted in noisy or dissipative environment. For this reason, it is important to find a way to evaluate, at least approximately, how our generation protocol is affected by such irreversible effects. A complete treatment of the problem would involve modeling the qubit decoherence and relaxation as well as different dissipative effects in the nanomechanical resonators. It is not our intention here to account for all these noise mechanisms. Instead, we will present one simple  situation which allows of a very illustrative \emph{exact solution}. We consider the case in which both NEMS (with lowering operators $a$ and $b$) lose energy to their surrounding with decay rates $\kappa_a$ and $\kappa_b$, respectively. For simplicity, we will not include the qubit decoherence and relaxation. This is justified if the qubit decoherence times are longer compared to the resonators ones. At present, the charge qubits are notably more robust against decoherence and relaxation than the nanomechanical resonators.  In this situation, and considering $\lambda_1=\lambda_2$, the system master equation at zero temperature, when expressed in terms of the mode operators, is written as
\begin{eqnarray}
\frac{\partial\tilde{\rho}_I}{\partial t}&=&-i[\mu(a_1^\dag a_1)^2\sigma_x,\tilde{\rho}_I]\nonumber \\
& & 
+\frac{\kappa_a+\kappa_b}{4}(2a_1\tilde{\rho}_Ia_1^{\dag}-a_1^{\dag}\hat{a_1}\tilde{\rho}_I-\tilde{\rho}_Ia_1^{\dag}a_1)\nonumber\\ &&+\frac{\kappa_a+\kappa_b}{4}(2a_2\tilde{\rho}_Ia_2^{\dag}-a_2^{\dag}\hat{a_2}\tilde{\rho}_I-\tilde{\rho}_Ia_2^{\dag}a_2)-
\nonumber \\
& & 
\frac{\kappa_a-\kappa_b}{4}[(a_1^\dag a_2+a_2^\dag a_1)\tilde{\rho}_I+\tilde{\rho}_I(a_1^\dag a_2+a_2^\dag a_1)]\nonumber\\ &&+\frac{\kappa_a-\kappa_b}{4}[2(a_1+a_2)\tilde{\rho}_I(a_1^\dag+a_2^\dag)].\label{me}
\end{eqnarray}
We can see that the master equation contains extra terms due to the transformation to normal modes. The full treatment for arbitrary $\kappa_a$ and $\kappa_b$ makes analytical progress quite difficult \cite{arthur} and a numerical calculation may be presented elsewhere. However, the simple regime in which both resonators decay with similar rates can be readily investigated. In this case, $|\kappa_a+\kappa_b|\gg|\kappa_a-\kappa_b|$, and we can drop the terms proportional to $(\kappa_a-\kappa_b)$. This assumption is realistic here since both resonators are assumed to be identical (same mass and natural frequencies). Even when this is not exactly the case, the quantities calculated below under the assumption of $(\kappa_a\approx\kappa_b)$ will, at least, serve as an upper bound to the case in which the dissipation rates are disparate.

\begin{figure}[h]
\begin{center}
\includegraphics[scale=1]{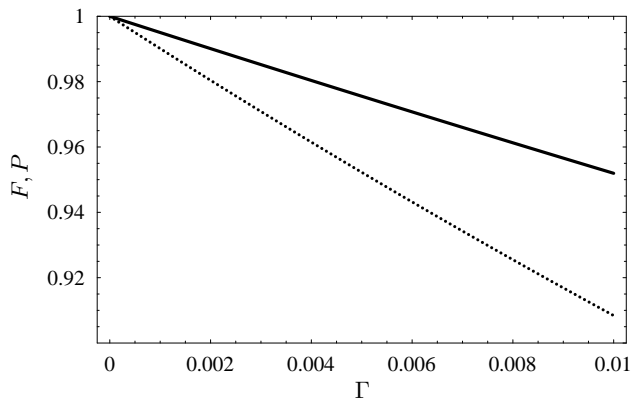}
\end{center}
\caption{Fidelity (solid) and purity (dotted) in function of the dimensionless parameter $\Gamma=\kappa/\mu$ for $\alpha=2$, at the moment the Yurke-Stoler state would be perfectly generated in the ideal lossless case.}
\label{fig2}
\end{figure}

We calculate here the degree of purity and the fidelity of the state, generated in such a noisy environment, as compared to the Yurke-Stoler state obtained in the ideal unitary case. Therefore, we need to take the same initial preparation used in the ideal case i.e. $|\psi(0)\rangle=|\alpha\rangle_a|\alpha\rangle_b|+\rangle_x$. As this implies that mode-$2$ will be in the vacuum state,  we need  to consider only the terms in (\ref{me}) that contain operators for mode-$1$. The master equation (\ref{me}) reduces to,
\begin{eqnarray}
\frac{\partial\tilde{\rho}_I}{\partial t}&=& -i[\mu(a_1^\dag a_1)^2,\tilde{\rho}_I]+\kappa(2a_1\tilde{\rho}_Ia_1^{\dag}-a_1^{\dag}\hat{a_1}\tilde{\rho}_I \nonumber \\
& & -\tilde{\rho}_Ia_1^{\dag}a_1), \label{mes}
\end{eqnarray}
where $\kappa=(\kappa_a+\kappa_b)/4$. An exact solution for (\ref{mes}) using the Q-function approach is presented in \cite{dissip}, but we will use a recent solution  obtained directly for the density operator \cite{hector} which allows us to  readily obtain the purity $P={\rm{Tr}}[\rho^2]$, and fidelity $F=\langle {\rm{YS}}|\rho|{\rm{YS}}\rangle$. According to \cite{hector}, the solution of (\ref{mes}) is  
\begin{widetext}
\begin{eqnarray}
\tilde{\rho}_I(t)&=& \left\{ \sum_{k,n,m=0}^{\infty}
\tilde{\rho}_{n+k,m+k}(0)e^{-i\mu t(n^2-m^2)-\kappa t(n+m)}\sqrt{\frac{(n+k)!(m+k)!}{n!m!}} \left[\frac{1-e^{-2i\mu t(n-m)-2\kappa t}}{2i\mu (n-m)+2\kappa}\right]^k  \frac{(2\kappa)^k}{k!}|n\rangle_1 {}_1\langle m| \right\}\nonumber\\ && \otimes|0\rangle_2{}_2\langle 0|\otimes|+\rangle_x {}_x\langle +|,
\end{eqnarray}
\end{widetext}
where $\tilde{\rho}_{n,m}(0)$ are the (Fock) matrix elements of the initial density matrix of mode-1. In Fig.(\ref{fig2}), we show the decay of fidelity (solid) and purity (dotted), as a function of the dimensionless parameter $\Gamma=\kappa/\mu$ for $\alpha=2$, at the time for which the Yurke-Stoler state arises, i.e., $\mu t_I=\pi/2$. We can see that the fidelity is quite high  ($F>0.99$) for $\Gamma\leq 10^{-3}$. As expected, Fig.{\ref{fig2}} reveals that the purity is more affected with increasing $\Gamma$ than is the fidelity. However, it also presents satisfactory values for $\Gamma\approx 10^{-3}$ ($P\approx 0.99$). For more realistic values such as $\Gamma\approx 10^{-2}$, we find $F>0.95$ and $P>0.90$.  

Finally, a few words about detection of superpositions of pairs of distinguishable coherent states is in order. This is an important topic, and several methods for detecting these states have already been proposed in the literature \cite{detdav,detwin,det}. Among them, it seems that the most suitable method for the system treated here is the one presented in \cite{detwin}, whereby motional states of a single trapped ion have been experimentally determined. This method relies upon implementation of displacement operators and Jaynes-Cummings interactions to determine both the density matrix in the number state basis and the Wigner function. Thi ion techniques could be an alternative to detect the Yurke-Stoler state proposed in this paper, but it should be remarked that a CPB coupled to two NEMS has not yet been operated in strong coupling regime.                   
\section{Conclusions}
\label{sec4}
To summarize, we have proposed a theoretical scheme to engineer a nonlinear Kerr Hamiltonian using superconducting charge qubits and nanoresonators. We have shown how such systems may be used to mimic a Kerr Hamiltonian. The formation of the Yurke-Stoler states in the composite mode of both resonators occurs naturally at an appropriate interaction time without needing to make a measurement on the system. For the case in which both resonators have equal decay rates, a simple exact expression for the total density matrix was derived. The present treatment, while not complete (more complex models of dissipation and noise could be considered), serves as an upper bound for the case in which the qubit decoherence can be neglected. In this context, we have shown that the fidelity of the generated state can high for moderate values of the decay constants. As a final remark, recently Woolley \emph{et al.} \cite{woolley2} have proposed a new protocol for high precision measurement in a nanomechanical resonator that makes explicit use of a Kerr nonlinearity. The method of the present paper could enable the use of linear nanomechanical resonators for such measurements instead of the intrinsically nonlinear nanomechanical resonators assumed in \cite{woolley2}.

\emph{Acknowledgments:} FLS wishes to thanks F. Brito for helpful discussions
and KF wishes to thank the Australian Research Council Centre for Quantum Computer Technology at The University of Queensland for hosting her visit to Brisbane. GJM would like to acknowledge the support of the Australian Research Council, FLS to FAPESP (Brazil) and CNPq (Brazil), and KF to CNPq (Brazil).


\end{document}